# A Stochastic Hybrid Framework for Driver Behavior Modeling Based on Hierarchical Dirichlet Process

Hossein Nourkhiz Mahjoub
Networked Systems Lab
ECE Department
University of Central Florida
Orlando, FL, US
hnmahjoub@knights.ucf.edu

Behrad Toghi
Networked Systems Lab
ECE Department
University of Central Florida
Orlando, FL, US
toghi@knights.ucf.edu

Yaser P. Fallah
Networked Systems Lab
ECE Department
University of Central Florida
Orlando, FL, US
yaser.fallah@ucf.edu

*Abstract*— **Scalability is one of the major issues for real-world Vehicle-to-Vehicle network realization. To tackle this challenge, a stochastic hybrid modeling framework based on a non-parametric Bayesian inference method, i.e., hierarchical Dirichlet process (HDP), is investigated in this paper. This framework is able to jointly model driver/vehicle behavior through forecasting the vehicle dynamical time-series. This modeling framework could be merged with the notion of model-based information networking, which is recently proposed in the vehicular literature, to overcome the scalability challenges in dense vehicular networks via broadcasting the behavioral models instead of raw information dissemination. This modeling approach has been applied on several scenarios from the realistic Safety Pilot Model Deployment (SPMD) driving data set and the results show a higher performance of this model in comparison with the zero-hold method as the baseline.**

## I. INTRODUCTION

From a macro-system perspective, situational awareness is an imperative for any distributed dynamic system composed of interactive agents (nodes). This crucial feature, attained by the virtue of sensory information and communication amongst nodes, allows individual agents to keep the track of the overall system behavior and assists them to perform the necessary coordinated actions more properly. As an illustration, in a distributed vehicular safety system each vehicle needs to have a precise understanding of its proximity within a range of at least 300 meters, based on National Highway Traffic Safety Administration (NHTSA) technical reports [1]-[2]. This requirement is mandatory for critical Cooperative Vehicular Safety (CVS) applications to properly detect the dangerous and critical situations and perform appropriate reactions in a timely manner. These reactions could range from issuing warnings to the driver to taking the control of the vehicle in full autonomy case.

Inter-vehicle communication is essential to improve and extend the situational awareness, since the sensory data could be restricted and cover a limited range due to the existence of obstacles or other confining circumstances forced by the environment, such as dimness, fog, rain, etc. Dedicated Short Range Communication (DSRC) [3], is one of the current primary communication technologies for the vehicular ad-hoc networks (VANETs) and is highly expected to be mandated by US officials in the near future. This will require all automotive original equipment manufacturers (OEMs) to deploy DSRC communication devices in their brand new productions. Situational awareness could be attained in the DSRC-based VANETs by sharing every agent's (vehicle's) raw dynamic information, e.g. GPS (latitude, longitude), velocity, acceleration, heading, etc., enclosed in specific application layer messages known as Basic Safety Messages (BSMs). The definite BSM content format is defined by a message set dictionary under the SAE J2735 standard [4].

The innovative idea of extracting a predictive abstract model for vehicles' dynamics to be disseminated over the network, which has been initially proposed in [5] as the model-based communication (MBC) methodology and then investigated with more depth in [6], is immensely inspired by considering the serious inherent channel utilization constraints of available VANET communications technologies, e.g. DSRC. Various congestion control methods have been proposed so far in the literature [7]-[15], to mitigate the effect of inefficient DSRC channel utilization. Higher communications quality achieved by more efficient channel utilization in turn increases the overall system performance in different aspects such as behavior of critical safety applications and providing more room to send lower priority information, to name a few. Some of these communication strategies are currently regarded as the state of the art and nominated as the core of SAE J2945/1 congestion control standard [15], [16]. These approaches provide fascinating improvements in channel utilization in comparison with baseline DSRC communication policy, i.e. constant frequency broadcast of raw data BSMs, by continuously adapting different flexible parameters such as content, length or dissemination rate of the information packet to the wireless channel quality.

However, there is still a significant improvement potential by shifting the paradigm from any scheme of raw data communication to the model-based information networking as proposed in [5], [6]. In addition to the noticeably increased

This material is based on work supported by the National Science Foundation under CAREER Grant 1664968.

efficiency in channel bandwidth utilization, other significant advantage of applying MBC methodology versus the raw data communication is its capability to substantially increase the forecasting accuracy over the long prediction durations. This is due to the flexibility of this approach to update the model structure and/or parameters at the subject agent on the fly and updating the remote agents' knowledge of the updated model accordingly in a real-time fashion.

Appropriate scalable model derivation strategies capable of capturing high level driving patterns, such as what has been adopted and employed in this work, i.e. Switching Linear Dynamical Systems-Hierarchical Dirichlet Process- Hidden Markov Model (SLDS-HDP-HMM), in conjunction with a precise MBC-customized communication policy design could profoundly outperform the conventional forecasting schemes at the receiver (remote) agents. In the conventional approach remote agents always presume a predefined behavior (or roughly speaking a predefined model) of the subject agent, e.g. constant velocity/acceleration models, with no structural model updates. However, VANETs are composed of highly dynamic agents which need to be traced precisely to realize the aforementioned concept of situational awareness. Thus, this weak assumption of the subject agent's dynamics, which almost neglects the plausible evolution of its behavioral structure over time, results in a notable worse prediction quality and consequently lower situational awareness level in comparison with the MBC approach. In this work, adopting the notion of MBC, applicability of Stochastic Hybrid Systems (SHS)-based modeling schemes with underlying Markovian Switching Processes (MSP), specifically SLDS-HMMs, is explored and their forecasting precision of joint driver-vehicle behaviors is investigated. The underlying MSP describes the latent behavioral mode (state) changes inferred through observable information provided via Controller Area Network (CAN) of the host (subject) agent. Since new unforeseen behaviors might always be revealed by the human driver, the model predictions should theoretically be drawn from an infinite size sample set in order to support the theoretical infinite structural cardinality of the SHS-HMM model. To this end, a non-parametric Bayesian approach has been adopted and studied in this work, enabling the model to infinitely generate new behavioral modes if none of the already generated modes could adequately mimic the driver current behavior.

The rest of this paper is organized as follows. Mathematical structure of the adopted modeling framework is presented and discussed in details in section II. The forecasting performance of this model is then analyzed and evaluated over a set of realistic driving scenarios and its accuracy gain is presented in section III. Finally, section IV concludes the paper.

## II. Problem Statement

### A. Stochastic Hybrid System (SHS)-Based Model Derivation Framework

Markovian switching processes (MSPs) are capable of capturing the evolution trend of an observation sequence if the total effect of the sequence history on the next unobserved value is assumed to be encapsulated in and expressed by some finite-length set of the last observed points. More specifically, if $Y = \{y_1, y_2, ...\}$ denotes a sequence of observations with $r^{th}$ order Markovian property, then

$$P(y_n \mid \{y_1, y_2, ..., y_{n-1}\}) = P(y_n \mid \{y_{n-r}, y_{n-(r-1)}, ..., y_{n-1}\}) \quad (1)$$

where $P$ here denotes the conditional emission probability of $y_n$ given the observation history sequence.

However, for numerous real world processes, the successive values of an observable parameter do not directly fulfill the Markovian property. In those cases, they are assumed to be emitted from some Hidden states, each has a certain emission probability distribution. In hidden Markov models [17], the transition probability distributions are also defined among the hidden layer states. The most challenging HMM problem, namely parameter estimation, is inferring the most meaningful model parameters of the underlying states, i.e. their emission and transition distributions, from the observation sequence. This problem has been tackled in the literature by different methods, mainly iterative likelihood maximization schemes using expectation maximization (EM).

Although HMM is much more effective than conventional Markovian models to comprehend and reveal the actual dependencies among observed sequence elements, it puts a strict limiting assumption on these observations. More specifically, HMM assumes that the observations are *independently* drawn from the emission distributions of the hidden states. Therefore, this model assumes observations as independent random variables, given the hidden state is known at each moment. This restrictive assumption forces the model to neglect the temporal dependencies between consecutive observed values. Combining the notion of hybrid system modeling with HMM framework is a promising candidate to address this limitation and take the temporal effects of the sequence history on its upcoming values into account. For instance, if the consecutive drawn observations of an HMM model are assumed to follow a certain dynamic model, e.g. an autoregressive (AR) model, within each hidden state, the resultant Stochastic Hybrid System (SHS) framework is able to perceive both inherent Markovian and temporal dependencies of the observations, simultaneously. This specific hybrid framework which is realized by switching among different dynamical behaviors based on the rules designated by an HMM model is known as Switching Linear Dynamical Systems (SLDS)-HMM model [18], [19]. Details of the adopted framework in this work which has been built upon the SLDS-HMM notion is described in the following section.

### B. Non-Parametric Bayesian SHS Formulation of Systems with Theoretically Infinite Behavioral Modes

In order to formulate a stochastic hybrid system, a SLDS-HMM here, with adaptive structural properties to the observations received from the target agent in an online manner, the model should be able to continuously track the agent behavior and add/remove the necessary/unnecessary behavioral modes on the fly. What is meant by structural properties here are the number of inferred discrete modes (states), dynamic behaviors assigned to each mode, etc. Therefore, the model is theoretically expected to generate infinite number of modes, since we do not want to put any restrictive assumptions on the model cardinality (number of states) and the dynamic behaviors defined by different modes. To this end, a nonparametric Bayesian approach, based on hierarchical Dirichlet process [20] is nominated and its performance to model a highly dynamic system, i.e. VANET, is investigated in this work. This extension of SLDS-HMM, known as SLDS-HDP-HMM, proposed in [19], leverages the properties of Dirichlet processes to define the transition probability measures among unbounded and unknown number of discrete modes. Very few previous studies, such as [21], have also studied non-parametric approaches to study driving behaviors. However, their studies are mainly focused on HDP-HMM models based on Beta process which is conceptually different from what we have in this work, i.e. SLDS-HDP-HMM model.

Any single draw from a Dirichlet process with a base measure $H$ and concentration parameter $\lambda$ , $DP(\lambda, H)$ ,is composed of an infinite number of $\Theta_i$s, so that $\Theta_i \in \Theta, \forall i = 1,2, \dots$ . Here $\Theta$ specifies the parameter space of the base measure, $H$.

However, since any two different sets of draws from a continuous distribution are fully disjoint and have definitely no overlap, taking the transition probability measures at different points of time as independent draws of a Dirichlet process with a continuous base measure forces the set of inferred states after each observation to be completely disjoint from all other sets of the already visited (or drawn) states. In other words, with a continuous base measure a visited behavioral mode of the system will never be revisited. To handle this issue, a discrete probability measure with the infinite capability of generating a new unobserved state with non-zero probability, i.e., a discrete measure with countably infinite parameter domain, should be selected as the base measure. Assigning another Dirichlet process for this purpose solves the problem and results in the hierarchical Dirichlet process-HMM structure.

In this work we have considered the sticky version of HDP-HMM [22]. The sticky property notes another extension to the idea of HDP-HMM and allows capturing modal behavior of the system in a more accurate manner. This is crucial to model the systems with persistent dynamic modes and prevent the model from rapid fluctuations among inherently similar modes.

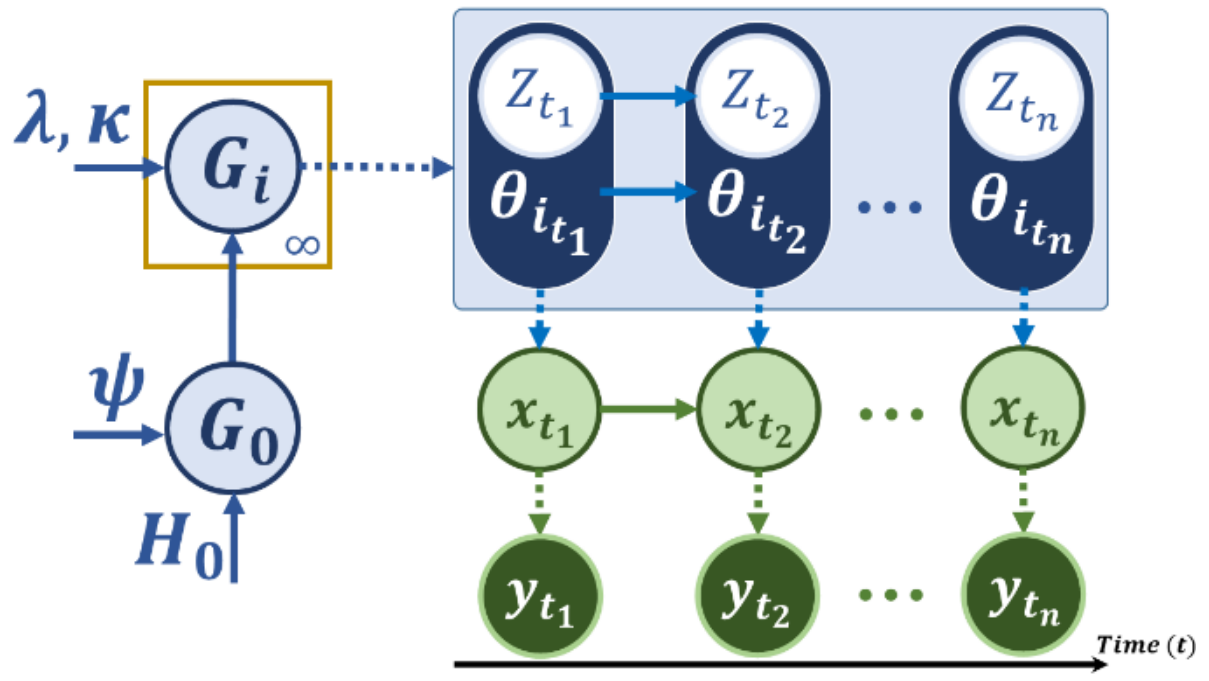

Figure 1. Generative model of hierarchical Dirichlet process

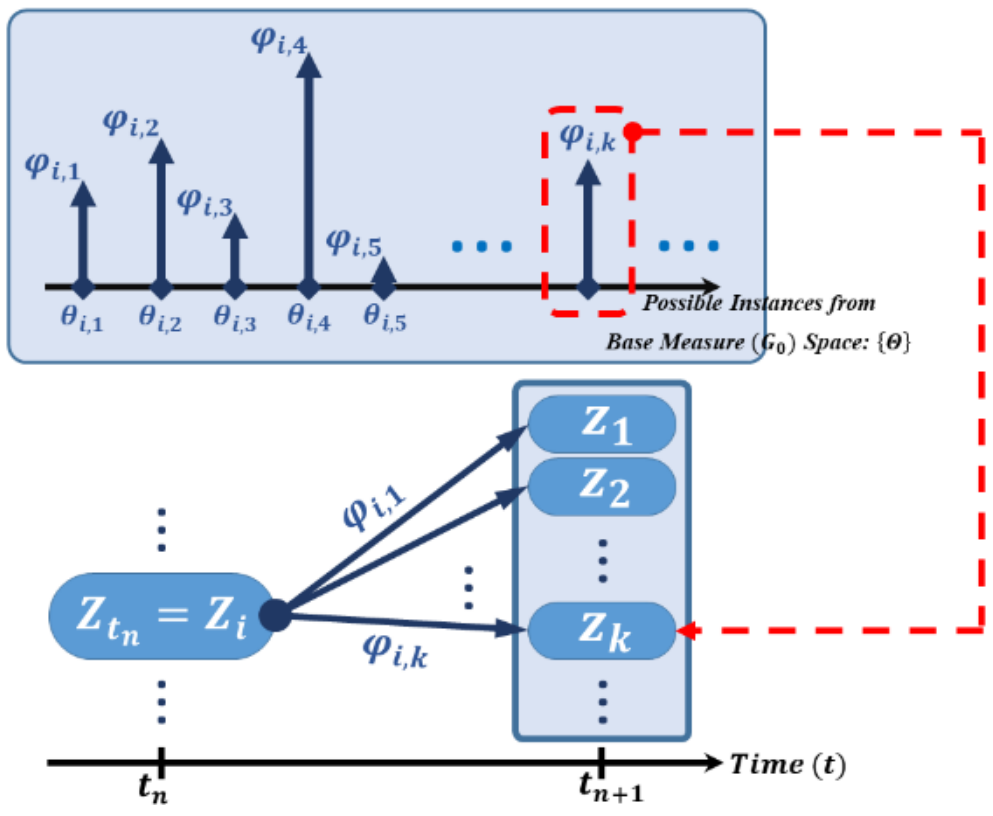

Figure 2. Evolution of a HDP-HMM mode structure over time. $\boldsymbol{G_i}$ is the transition probability measure of the $\boldsymbol{i^{th}}$state ($\boldsymbol{Z_i}$), and $\boldsymbol{Z_{t_n}} \in \{\boldsymbol{Z_1}, \boldsymbol{Z_2}, \dots\}$ denotes the hidden state at the time $\boldsymbol{t_n}$. It should be noted that: $\boldsymbol{Z_i \sim G_i, G_i \sim DP(\lambda, G_0), G_0 \sim DP(\psi, H_0)}$.

Generative model of SLDS-HDP-HMM and its evolution over time are depicted in Figure 1 and Figure 2, respectively. In these figures, observations, discrete and continuous states at time instant $t_n$ , are denoted by $y_{t_n}$ , $z_{t_n}$ , and $x_{t_n}$ , respectively. Discrete states are what have been also referred to as behavioral modes so far. These parameters form a state space model within an SLDS structure:

$$x_{t_n} = A^{(z_{t_n})} x_{t_{n-1}} + e(z_{t_n}) \tag{2}$$

$$y_{t_n} = C x_{t_n} + w(t_n) \tag{3}$$

Where $A$ and $e$ are process matrix and process noise parameters respectively, derived from a mode-specific distribution. In addition, $C$ and $w$ are mode independent (could also be mode-specific) parameters representing the measurement matrix and observation noise distribution, respectively. The mode specific distributions for $A$ and $e$ are from a parameter space $\Theta$ . Therefore, the transition probability measures, denoted as $G_i$ (in a sum of weighted unit mass function form), could be drawn from a Dirichlet process which its base measure has the same parameter space $\Theta$. The weights of this Dirichlet process are elements of a draw from a Griffith-Engen-McCloskey (GEM) distribution . GEM distribution, which is a distribution over the countably infinite size probability measures ($\Phi$s), has a generative model, known as stick-breaking construction, which guarantees that the weights ($\varphi_k$ probabilities) always add up to 1. Utilizing this specific generative model allows drawing state transition probabilities without knowing the exact number of states in advance.

Since any Dirichlet process is theoretically an infinite size stochastic measure, it requires a practical generative model in order to be realized and implemented. Chinese Restaurant Process (CRP) is a Pólya urn-based process which allows to generate a Dirichlet process without developing its complete realization. Tracing back, the parameters of the DP and each transition probability is computed as:

$$\nu_k | \lambda \sim Beta(1, \lambda) \qquad for\ k = 1, 2, \dots \tag{4}$$

$$\varphi_k = \nu_k \prod_{h=1}^{k-1} (1 - \nu_h) \qquad for\ k = 1, 2, \dots \tag{5}$$

$$\Phi \sim GEM(\lambda) \iff \Phi = \{\varphi_1, \varphi_2, \dots\},\ \varphi_k \in (0,1),\ (\forall k = 1, 2, \dots),\ \sum_{k=1}^{\infty} \varphi_k = 1 \quad (6)$$

$$G_i \sim DP(\lambda, G_0) \iff G_i = \sum_{k=1}^{\infty} \varphi_{i,k} \delta_{\theta_{i,k}}$$
$$\left\{ \begin{matrix} \theta_{i,k} | G_0 \sim G_0 \\ \{\varphi_{i,:}\} \sim GEM(\lambda) \end{matrix} \right\} for\ k = 1, 2, \dots \quad (7)$$

$$G_0 \sim DP(\psi, H_0) \quad (8)$$

The mentioned stick-breaking model is realized by the means of a Beta distribution with parameters 1 and $\lambda$, as depicted in (4), and (5).

The complexity of the model generated by the virtue of a Dirichlet process is driven by and adapted to the history of the observations in an online manner. To this end, after receiving each new observation element Dirichlet process should decide how to reflect it into the model. This task is performed through the inherent CRP mechanism of the Dirichlet process. The probability of assigning the most recent data point to any of the already generated states is proportional to the number of observation elements which have been assigned to that state so far and is inversely proportional to the whole observation sequence length. This imitates the well-known effect of "rich get richer" in the model structure. Generating a new state for the most recent observation is proportional to the CRP parameter and is again inversely proportional to the whole observation sequence length. Thus, the probability of generating a new state tends to zero when the number of observations goes towards infinity. This could be interpreted as a monotonically increase in the accuracy of the inferred model by the Dirichlet process and approaching towards the complete behavioral model of the system through receiving more and more observations. After that the decision for the model state cardinality is made, the problem reduces to a normal HMM with a known number of modes. "Evaluation, Decoding, and Parameter Estimation" problems of this HMM now could be tackled utilizing well-known approaches in the literature, i.e. forward-backward algorithm, Viterbi algorithm, and Expectation-Maximization method, respectively.

## III. Evaluation

Derived SLDS-HDP-HMM model from the previous section is applied on an extensive driving information set of selected trips from a realistic rich driving data set provided by US DOT, namely Safety Pilot Model Deployment (SPMD) data set [23]. SPMD data set which is composed of information collected through two different settings of Data Acquisition Systems (DAS1 and DAS2) in Ann-Arbor, Michigan, provides different in-vehicle information logged from CAN, such as longitudinal velocity and acceleration, yaw rate, steering angle, turn flash status, etc., along with the vehicle positioning information over the whole trip duration. The data are analyzed in this work to realize the number of hidden driving behavioral states inferred by the model in real driving situations. In addition, the obtained model is employed to predict the future trend of the analyzed time series under imperfect network conditions. Network imperfection has been modeled through the Packet Error Rate (PER) abstraction, which is a common approach to consider wireless channel condition from the application layer point of view.

The sequence of hidden modes assigned by the model to a sample interval of the longitudinal acceleration is depicted in Figure 3. Figure 4 represents the predicted longitudinal speed of the same trip and its comparison to the baseline model, which is basically a zero-hold estimation model here. This comparison is performed under 60% PER rate and evidently demonstrates the forecasting dominance of our model during a notable portion of the scenario. The evaluated Empirical Cumulative Distribution Function (ECDF) of the prediction errors over 10 trips for both baseline and SLDS-HDP-HMM models are depicted in Figure 5, which shows higher prediction precision of our model for errors less than 1 meter.

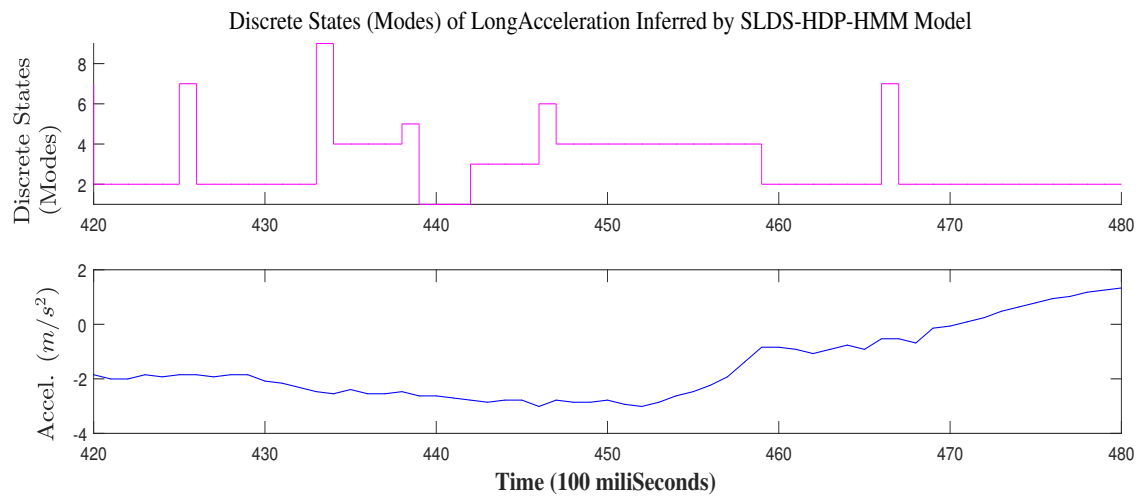

Figure 3. Discrete states (modes) of the longitudinal acceleration inferred by SLDS-HDP-HMM model

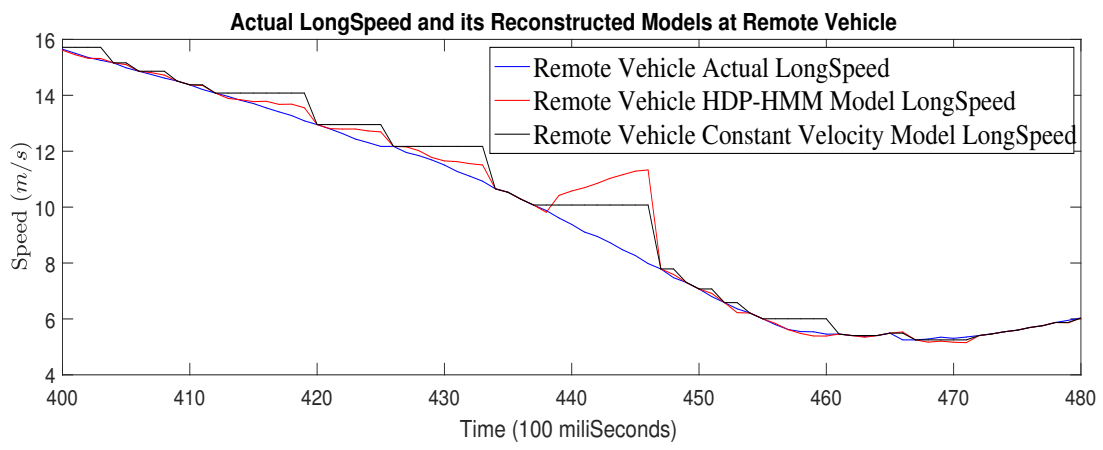

Figure 4. Prediction accuracy comparison of SLDS-HDP-HMM and Baseline models for longitudinal speed under 60% PER

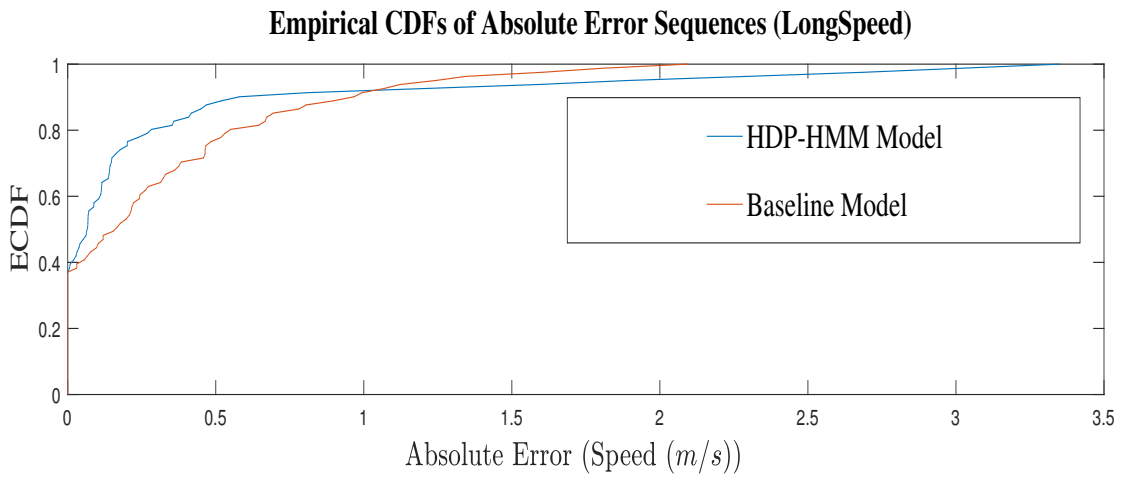

Figure 5. ECDF of prediction errors for longitudinal speed in SLDS-HDP-HMM and Baseline models under 60% PER

## IV. Conclusion

In this work a non-parametric Bayesian approach is investigated to track the joint vehicle-driver behavior in an online manner. A stochastic hybrid model is designed based on a hierarchical Dirichlet process. The HDP serves as the underlying Markovian switching process of the model and determines the appropriate sequence of hidden driving behavioral modes (states). The non-parametric nature of the employed modeling framework makes it possible to add an unbounded number of unforeseen states to the model on the fly. This is crucial due to the stochasticity of the system states (driving modes) generated by the human driver. The recorded in-vehicle CAN time series available in the realistic SPMD driving data set are employed for model derivation here. The results show a notable prediction performance improvement against the zero-hold model as the baseline using the SLDS-HDP-HMM framework.